\documentclass[prd,aps]{revtex4-1}

\usepackage{epsfig}
\usepackage{graphicx,color}

\newcommand{\beq}{\begin{equation}}
\newcommand{\eeq}{\end{equation}}
\newcommand{\be}{\begin{eqnarray}}
\newcommand{\ee}{\end{eqnarray}}

\begin{document}


\title{
$B_s \to \mu \tau$ and $h \to \mu \tau$ decays 
in the general two Higgs doublet model
}

\author{ Jong-Phil Lee }

\affiliation{
Sang-Huh College, Konkuk University, Seoul 05029, Korea
}

\author{ Kang Young Lee }
\email{kylee.phys@gnu.ac.kr}

\affiliation{
Department of Physics Education \&
Research Institute of Natural Science,
\\
Gyeongsang National University, Jinju 52828, Korea
}

\date{\today}

\begin{abstract}
 \noindent
Inspired by the recent measurement 
of the $h \to \mu \tau$ decays 
by the CMS collaboration at the LHC,
we study the lepton flavour-violating (LFV) 
$B_s \to \mu \tau$ decays
in the general two Higgs doublet model. 
Those LFV interactions could accommodate
the present deviation of the muon anomalous magnetic moment
and also predict the LFV $\tau$ decay processes 
such as $\tau \to \mu \mu \mu$ and $\tau \to \mu \gamma$.
We find that the $B_s \to \mu \tau$ decay rates
should be sizable with above experimental conditions
in the framework of our model. 
These processes are expected to be observed at the colliders 
such as LHCb and Belle-II in the future.
\\
\end{abstract}

\pacs{ }

\maketitle

\section{Introduction}
\label{sec:intro}

The discovery of a Higgs boson 
\cite{higgsATLAS,higgsCMS}
has opened a new era of particle physics.
Henceforth we have to explore the properties of this new boson in detail
and try to understand the whole structure of the Higgs sector.
Recently the CMS collaboration has reported a slight excess 
of an exotic decay mode of the Higgs boson into the $\mu \tau$
final states \cite{cms}.
The best fit value of the branching ratio is 
Br$(h \to \mu \tau)=(0.84^{+0.39}_{-0.37})$ \%
which shows a 2.4-$\sigma$ deviation 
from the null result predicted in the standard model (SM).
The measurement of the ATLAS collaboration also shows a deviation but
still less significance than the CMS result, 
Br$(h \to \mu \tau)=(0.77 \pm 0.62)$ \% \cite{atlas}.
The combined result is given by
\be
{\rm Br}(h \to \mu \tau)=(0.82^{+0.33}_{-0.33}) \%
\ee
and presents a upper limit to be 1.39 \% at 95\% C.L..

Since $h \to \mu \tau$ decays are 
the lepton flavour-violating (LFV) processes and 
forbidden in the SM, 
the excess could be a direct evidence 
of the new physics (NP) beyond the SM 
if it will be confirmed with more data in the future.
Lots of studies of the new physics explanation on 
the excess of Br$(h \to \mu \tau)$
has been provided in many literatures \cite{hmutau}.
In this letter we consider the general extension of the SM
with 2 Higgs doublets as a solution of the LFV Higgs decays.
The flavour-changing neutral current (FCNC) interactions
with scalars are generated at tree level
if the additional Higgs doublets exist
without some flavour conserving mechanism.
They lead to the LFV Higgs boson decays
in the general multi-Higgs doublets models.
The CMS collaboration has found no evidences
in the $h \to e \tau$ and $h \to e \mu$ channels
\cite{emuetau}.
Thus we focus on the scalar$-\mu-\tau$ couplings
and neglect other LFV interactions in this paper.

The new scalar$-\mu-\tau$ interactions 
provide various phenomenological implications.
First they generically contribute
to the muon anomalous magnetic moment, $(g-2)_\mu$.
The precise measurement of $(g-2)_\mu$ 
has been one of the most sensitive probe of the NP
and still shows unexplained deviation from the SM prediction 
more than 3-$\sigma$ at present
\cite{PDG}.
The scalar LFV interactions are helpful to accommodate the deviation
\cite{NieSher,KangLee}.
On the other hand, the LFV $\tau$ decays are also predicted 
with the scalar FCNC, while they are absent in the SM.
Thus the present experimental limits of $(g-2)_\mu$ 
and the LFV $\tau$ decays provide stringent constraints on the model.

Here we consider the LFV $B_s \to \mu \tau$ decays 
in the general two Higgs doublet model (2HDM).
The rare $B$ decay modes involving the FCNC are 
very good testing ground to find hints for NP
and have been studied in various channels.
For instance, the $B_s \to \mu^- \mu^+$ decays 
have been in the spotlight 
to explore the large supersymmetry contribution 
with scalar exchanges.
Recently the branching ratio of $B_s \to \mu^- \mu^+$ mode is measured 
by the LHCb and the CMS to be 
Br$(B_s \to \mu^- \mu^+)=(3.1 \pm 0.7) \times 10^{-9}$
\cite{bsmumulhcb,bsmumucms}
which agrees with the SM prediction.
We note that Br$(h \to \mu^- \mu^+)$ is of order 10 $\%$,
two order higher than the best fit value of Br$(h \to \mu \tau)$.
Assuming the SM Higgs mediated process is dominated 
in $B_s \to \mu \tau$ decays,
the ratios of $B_s \to \mu \tau$ to $B_s \to \mu \mu$ decays
are comparable with those of 
Br$(h \to \mu \tau)$ to Br$(h \to \mu \mu)$. 
Then we estimate the branching ratio of $B_s \to \mu \tau$
to be of order $10^{-11}$ and it is hard to be measured
in the near future. 
If there are additional contributions to $B_s \to \mu \tau$ decays,
however, its branching ratio might be large enough to be observed
while Br($B_s \to \mu \mu$) being kept to be within the present measurement
\cite{Bsmutau}.
We explore the possibility of such enhancement
including the other scalar contributions
in the general 2HDM framework.

The paper is organized as follows.
We briefly describe the lepton flavour-violation 
in the general two Higgs doublet model
and obtain the scalar$-\mu-\tau$ couplings
from the experiments including 
the $h \to \mu \tau$ decays measured at the LHC 
in Sec.~\ref{sec:model}. 
In Sec.~\ref{sec:tau},
we consider the muon anomalous magnetic moment $(g-2)_\mu$ 
and the LFV $\tau$ decay processes in this model.
In Sec.~\ref{sec:Bs},
the $B_s \to \mu \tau$ decays are studied
under the experimental constraints discussed in the previous sections.
Section~\ref{sec:concl} is devoted to conclusions.

\section{LFV in the general 2HDM}
\label{sec:model}

We can choose a basis for the two Higgs doublets 
$\hat{H}$ and $\hat{\Phi}$
where only one Higgs doublet $\hat{H}$ 
gets a vacuum expectation value (VEV)
and is responsible for the electroweak symmetry breaking
\cite{NieSher,davidson}.
After an appropriate rotation of leptons
such that the neutral components of $\hat{\Phi}$
has flavour-diagonal couplings,
the relevant Lagrangian for Yukawa interactions
of leptons and d-type quarks reads
\be
\cal{L} &=& \frac{\sqrt{2}}{v} \left(
  m_e \bar{e}_L e_R + 
  m_\mu \bar{\mu}_L \mu_R + 
  m_\tau \bar{\tau}_L \tau_R \right) H^0
 + h^l_{ij} \bar{l}_{iL} l_{jR} \phi^0 
\nonumber \\
&& + \frac{\sqrt{2}}{v} \left(
  m_d \bar{d}_L d_R + 
  m_s \bar{s}_L s_R + 
  m_b \bar{b}_L b_R \right) H^0
 + h^d_{ij} \bar{d}_{iL} d_{jR} \phi^0 
\label{eq:lagrangian}
\ee
where the neutral components consist of
\be
H^0 &=& \frac{1}{\sqrt{2}} \left(v + H_s +i G^0 \right),
\nonumber \\
\phi^0 &=& \frac{1}{\sqrt{2}} \left( \phi_s + i \phi_p \right) ,
\ee
with the scalars $H_s$ and $\phi_s$, Goldstone mode $G^0$, 
and the pseudoscalar $\phi_p$.
Assuming that the CP is conserved in the Higgs sector,
the physical states of CP-even scalars, 
$h$ and $H$ are defined through the mixing
\be
\left(
  \begin{array}{c}
    \phi_s \\
    H_s \\
  \end{array}
\right) 
=
\left(
  \begin{array}{cc}
    \cos\theta & -\sin\theta \\
    \sin\theta & \cos\theta \\
  \end{array}
\right)
\left(
  \begin{array}{c}
    H \\
    h \\
  \end{array}
\right)~,
\ee
and the CP-odd scalar $A = \phi_p$.

The SM-like Higgs boson $h$ decays into the LFV final states
through the small mixing $\sin \theta$.
The decay width is given by
\be 
\Gamma (h \to \mu \tau) = \frac{m_h \sin^2 \theta}{16\pi} 
        \left( |h_{\mu \tau}|^2 + |h_{\tau \mu}|^2 \right) ,
\ee
and the corresponding branching fraction given by 
${\rm Br}(h \to \mu \tau) 
  = \Gamma (h \to \mu \tau)/(\Gamma_{SM} + \Gamma (h \to \mu \tau))$.
From now on the Yukawa couplings are assumed to be real 
and $h_{\mu \tau}=h_{\tau \mu}$ for simplicity. 
Thus we obtain the relation for the combined parameter 
$h_{\mu \tau} \sin \theta$,
\be
h_{\mu \tau}^2 \sin^2 \theta = \frac{8 \pi}{m_h} \Gamma_{SM}
          \frac{{\rm Br}(h \to \mu \tau)}{1-{\rm Br}(h \to \mu \tau)}
    \approx 0.68 \times 10^{-5} 
          \left( \frac{{\rm Br}(h \to \mu \tau)}{0.82 \%} \right)  .
\ee

Two curves in the Fig. 1 depicts the equation (6) at 95 $\%$ C.L.
on the plane of the mixing angle $\sin \theta$ 
and the LFV coupling $h_{\mu \tau}$. 
The region between two curves denotes the allowed values
by the $h \to \mu \tau$ branching ratio measurements.
We see that the mixing angle might be large 
if $h_{\mu \tau}$ is small enough as shown in the plot.

\begin{figure}
\begin{center}
\includegraphics[width=.6\textwidth]{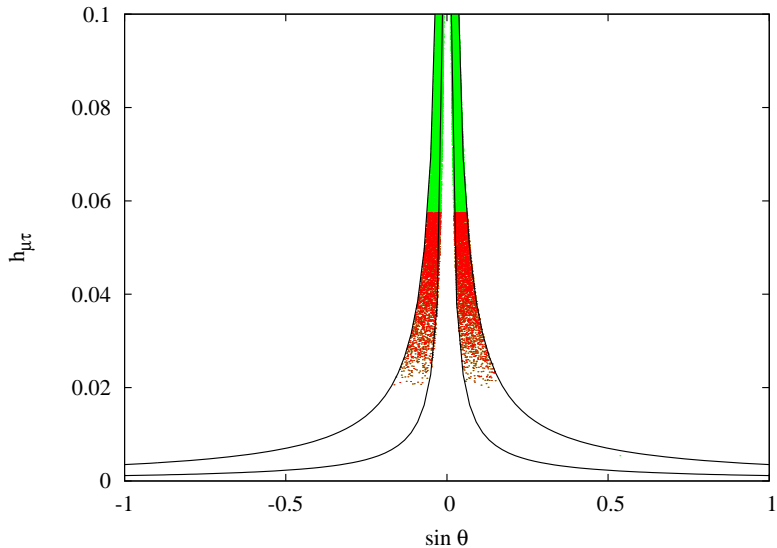}
\caption{
Allowed parameter sets of $(\sin \theta, h_{\mu \tau})$.
The region between curves denotes the parameters 
allowed by $h \to \mu \tau$ decays,
the green dots (overlapped by red dots) are constrained by $(g-2)_\mu$,
and the red dots additionally constrained 
by LFV $\tau \to \mu \gamma$ and $\tau \to \mu \mu \mu$.
}
\label{fig:f1}
\end{center}
\end{figure}

\section{ LFV constraints}

\subsection{ LFV contributions on $\Delta a_\mu$ }
\label{sec:amu}

The LFV scalar interactions also induce new contributions 
to the muon anomalous magnetic moment, $(g-2)_\mu$. 
Still the experimental data of $(g-2)_\mu$ 
shows a deviation more than 3-$\sigma$ from the SM prediction as,
\be
\Delta a_\mu \equiv a_\mu^{\rm exp} - a_\mu^{\rm SM}
          = (288 \pm 63 \pm 49) \times 10^{-11},
\ee
where the first error is experimental and the second theoretical. 
The LFV scalar interaction is one of the good candidates 
to cure this disagreement of $(g-2)_\mu$
between theory and experiments.
The leading contribution to the $(g-2)_\mu$ are given by
\be
\Delta a_\mu = \frac{h_{\mu \tau}^2}{16 \pi^2} m_\mu m_\tau
      \left[ 
         \frac{\sin^2 \theta}{m_h^2} 
         \left( \log \frac{m_h^2}{m_\tau^2}-\frac{3}{2} \right) 
        +\frac{\cos^2 \theta}{m_H^2} 
         \left( \log \frac{m_H^2}{m_\tau^2}-\frac{3}{2} \right) 
        -\frac{1}{m_A^2} 
         \left( \log \frac{m_A^2}{m_\tau^2}-\frac{3}{2} \right) 
      \right],
\ee
in the general 2HD model.
We note that the SM Higgs contribution of the first term in Eq. (8) 
$\sim 4.4 \times 10^{-12}$ with the value of Eq. (6),
which could not explain the deviation
and additional contribution of $H$ are inevitable to accommodate
$\Delta a_\mu$ in this model.
If the FCNC Yukawa couplings $h_{ij}$ are small enough,
$H$ and $A$ might be lighter than the SM Higgs boson $h$ 
in this general model.
However we avoid unnatural fine tuning and 
assume the conservative condition $m_H,m_A \ge m_h$ in this analysis.

We scan the model parameters 
$(\sin \theta$, $h_{\mu \tau}, m_H, m_A)$
with the constraints given in the previous section.
The red dots and green dots in the Fig. 1 are allowed values 
of $\sin \theta$ and $h_{\mu \tau}$
by $\Delta a_\mu$ data at 95 $\%$ C. L.. 
The large mixing angle regions are excluded 
and $| \sin \theta | < 0.16$.
Note that the negative contribution of $A$ 
cancels the $H$ and $h$ contributions in Eq. (8)
and large LFV coupling $h_{\mu \tau} \sim 0.1$ is still allowed.

\subsection{LFV $\tau$ decays }
\label{sec:tau}

The Higgs FCNC couplings lead to the various LFV decay processes,
which do not exist in the SM.
In this letter, we focus only on the scalar$-\mu-\tau$ coupling
and the relevant LFV decays are 
$\tau \to \mu \gamma$ and $\tau \to \mu \mu \mu$.
The strong experimental limits are given by
Br$(\tau \to \mu \gamma) < 4.4 \times 10^{-8}$ 
and Br$(\tau \to \mu \mu \mu) < 2.1 \times 10^{-8}$ 
\cite{PDG}.

We write the effective lagrangian 
for electromagnetic penguin operators as
\be
{\cal L}_{\rm eff} = C_L {\cal O}_L + C_R {\cal O}_R + H.c.~~,
\ee
where the operators are given by
\be
{\cal O}_{L,R} = \frac{e}{8 \pi^2} m_\tau 
     \left( 
       \bar{\mu} \sigma^{\mu \nu} P_{L,R} \tau 
     \right) F_{\mu \nu}
\ee
and the leading contributions to the one-loop and two-loop 
Wilson coefficients by
\cite{harnik}
\be 
C_{L,R}^{(1)} &\approx& 
        \frac{1}{4 m_h^2} \frac{m_\tau}{v} h_{\mu \tau} \cos \theta
              \left(  \log \frac{m_h^2}{m_\tau^2} - \frac{4}{3} \right)
\nonumber \\
C_{L,R}^{(2)} &\approx& 0.055 h_{\mu \tau} \frac{1}{(125~{\rm GeV})^2}.
\ee 
Note that the one-loop contributions 
involve the $\tau$ internal line diagrams and 
the two-loop contributions come from the Barr-Zee type diagrams.
The branching ratio for $\tau \to \mu \gamma$ decay is
\be
{\rm Br}(\tau \to \mu \gamma) = \tau_\tau 
             \frac{\alpha m_\tau^5}{64 \pi^4} 
                   \left( |C_L|^2 + |C_R|^2 \right), 
\ee
where $\tau_\tau$ is the tau lifetime. 

Due to the Higgs LFV coupling,
the $\tau \to \mu \mu \mu$ decay is obtained at tree level 
through the Higgs mediated diagram.
The branching ratio for $\tau \to \mu \mu \mu$ decay is given by
\be
{\rm Br}(\tau \to \mu \mu \mu) = \tau_\tau 
\frac{m_\tau^5}{3072 \pi^3} h_{\mu \tau}^2
 \left( \left| \frac{\sin \theta}{m_h^2} y_{h\mu \mu}
       - \frac{\cos \theta}{m_H^2} y_{H\mu \mu} \right|^2
       + \left| \frac{1}{m_A^2} y_{A\mu \mu} \right|^2 \right),
\ee
where the lepton flavour conserving Higgs couplings are
\be
y_{h\mu \mu} &=& \frac{m_\mu}{v} \cos \theta 
                - \frac{h_{\mu \mu}}{\sqrt{2}} \sin \theta,
\nonumber \\
y_{H\mu \mu} &=& \frac{m_\mu}{v} \sin \theta 
                + \frac{h_{\mu \mu}}{\sqrt{2}} \cos \theta,
\nonumber \\
y_{A\mu \mu} &=& \frac{h_{\mu \mu}}{\sqrt{2}} ,
\ee
where the new flavour conserving coupling $h_{\mu \mu}$
is assumed to be the same order of the ordinary
Yukawa coupling $\sim m_\mu/v$ here.

The red dots in the Fig. 1 denotes the allowed values
of $\sin \theta$ and $h_{\mu \tau}$
by the additional constraints of the absence of 
$\tau \to \mu \gamma$ and $\tau \to \mu \mu \mu$ decays
at 95 $\%$ C. L..
We see that 
the limit of LFV $\tau \to \mu \gamma$ decay directly leads to
the upper bound on $h_{\mu \tau} \sim 0.06$
at this confidence level. 

We show the masses of extra neutral scalars $H$ and $A$
in Fig. 2 and 3 with the allowed values of
$\sin \theta$ and $h_{\mu \tau}$.
Since the sizable $H$ contribution is required 
to accommodate $\Delta a_\mu$ data, 
the $H$ mass is upper bounded depending upon $h_{\mu \tau}$ 
and has the absolute upper bound $\sim$ 420 GeV
as shown in Fig. 2.
No limits are attributed to the $A$ mass.
Moreover allowed is the parameter region
where both of the $H$ and $A$ are very light simultaneously
since the negative contribution of $A$ 
cancels the $H$ contribution in $\Delta a_\mu$ calculation.

\begin{figure}
\begin{center}
\includegraphics[width=.6\textwidth]{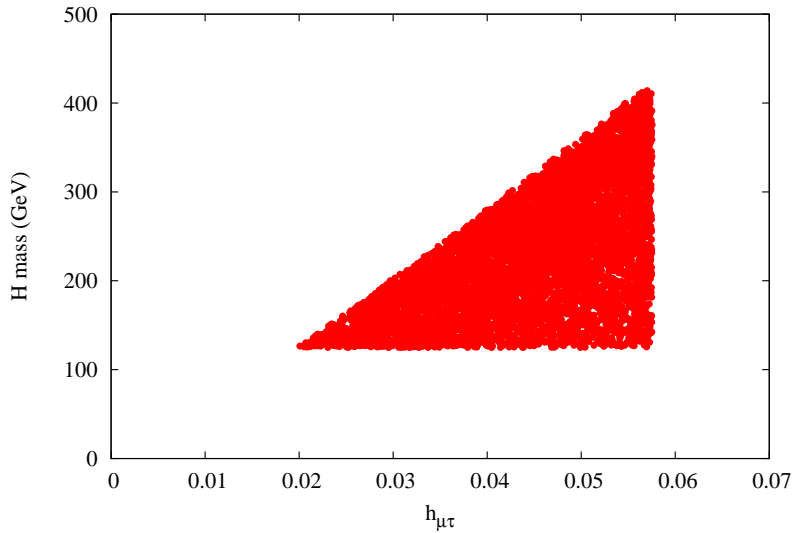}
\caption{
Allowed masses of $H$ with respect to $h_{\mu \tau}$ 
by $h \to \mu \tau$ decays, $(g-2)_\mu$, $\tau \to \mu \gamma$,
and $\tau \to \mu \mu \mu$ decays.
}
\label{fig:f2}
\end{center}
\end{figure}

\begin{figure}
\begin{center}
\includegraphics[width=.6\textwidth]{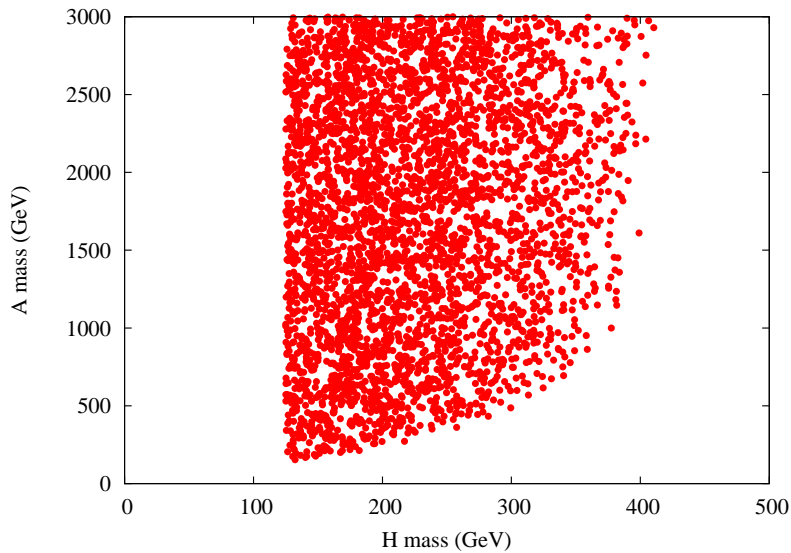}
\caption{
Allowed masses of $H$ and $A$
by $h \to \mu \tau$ decays, $(g-2)_\mu$, $\tau \to \mu \gamma$,
and $\tau \to \mu \mu \mu$ decays.
}
\label{fig:f3}
\end{center}
\end{figure}


\section{LFV $B_s \to \mu \tau$ Decays}
\label{sec:Bs}

Study of the $B_s$ phenomenology has been performed
at the Tevatron and becomes animated at the LHCb.
The $B_s$ meson provides good probes to the NP
since it involves relatively large FCNC interactions.

The relevant terms of the effective Hamiltonian for $B_s$ decays
contributing to the LFV decays of $B_s$ mesons are
\be
{\cal H}_{\rm eff} =  -\frac{G_F^2 M_W^2}{\pi^2} V_{tb}^* V_{ts} 
       \left( C_{10} {\cal O}_{10} + C_S {\cal O}_S + C_P {\cal O}_P \right)
       + H.c.~,
\ee
where the operators are given by
\be
{\cal O}_{10}&=&(\bar{b}_R \gamma^\mu s_L)
                (\bar{\mu} \gamma_\mu \gamma_5 \tau),
\nonumber \\
{\cal O}_S &=& m_b (\bar{b}_R s_L)(\bar{\mu} \tau),
\nonumber \\
{\cal O}_P &=& m_b (\bar{b}_R s_L)(\bar{\mu} \gamma_5 \tau).
\ee
The Wilson coefficients are obtained 
from the $h$, $H$, and $A$ exchange diagrams in this model,
\be
C_S 
&=& -\frac{\pi^2}{2 G_F^2 M_W^2 (V_{tb}^* V_{ts})}
            \frac{h_{bs} h_{\mu \tau}}{m_b}
          \left( \frac{\sin^2 \theta}{m_h^2} 
                 + \frac{\cos^2 \theta}{m_H^2} \right) ,
\nonumber \\
C_P 
 &=& \frac{\pi^2}{2 G_F^2 M_W^2 (V_{tb}^* V_{ts})}
            \frac{h_{bs} h_{\mu \tau}}{m_b}
            \frac{1}{m_A^2}. 
\ee
We also assume that $h_{bs}=h_{sb}$ and is real for simplicity.
Then the branching ratio of $B_s$ mesons are given by
\be
{\rm Br}(B_s \to \mu \tau) &=& \frac{G_F^4 M_W^4}{8 \pi^5}
             |V_{tb}^* V_{ts}|^2 M_{B_s}^5 f_{B_s}^2 \tau_{B_s} 
                  \left( \frac{m_b}{m_b+m_s} \right)^2 
\nonumber \\
   && \times \sqrt{ \left(1-\frac{(m_\tau+m_\mu)^2}{M_{B_s}^2}\right)
                    \left(1-\frac{(m_\tau-m_\mu)^2}{M_{B_s}^2}\right) }
\nonumber \\
   && \times \left[ \left(1-\frac{(m_\tau+m_\mu)^2}{M_{B_s}^2}\right) |C_S|^2
         + \left(1-\frac{(m_\tau-m_\mu)^2}{M_{B_s}^2}\right) |C_P|^2 \right].
\ee

The quark sector FCNC coupling $h_{bs}$ is constrained by the $B$ physics data.
We consider the $B_s-\bar{B}_s$ mixing as a constraint for $h_{bs}$.
The present measurement of the mass difference $\Delta M_s$ 
\cite{PDG}
\be
\Delta M_s =17.756 \pm 0.021 
\ee
in $10^{12} \hbar$ s$^{-1}$.
The $\Delta M_s$ in the general 2HDM reads
\cite{buras}
\be
\Delta M_s &=& \Delta M_s^{\rm SM} 
     + 2 h_{bs}^2 \left[ 
           \frac{\sin^2 \theta}{m_h^2} \Delta_h 
    + \frac{\cos^2 \theta}{m_H^2} \Delta_H
    - \frac{1}{m_A^2} \Delta_A
                     \right],
\ee
where 
\be
\Delta_S &=& \sum_{i=1,2} \left(
       C_{Si}^{SLL}(\mu) \langle O_i^{SLL}(\mu) \rangle
     + C_{Si}^{SRR}(\mu) \langle O_i^{SRR}(\mu) \rangle
     + C_{Si}^{LR}(\mu) \langle O_i^{LR}(\mu) \rangle \right),
\ee
with $S=h$, $H$, $A$. 
The Wilson coefficients up to ${\cal O}(\alpha_s)$ are
\be
C_{S1}^{SLL}(\mu) &=& C_{S1}^{SRR}(\mu) 
     = 1 + \frac{\alpha_s}{4\pi} 
     \left( -3 \log \frac{m_S^2}{\mu^2} + \frac{9}{2} \right)
\nonumber \\
C_{S2}^{SLL}(\mu) &=& C_{S2}^{SRR}(\mu) 
     =  \frac{\alpha_s}{4\pi} 
     \left( -\frac{1}{12} \log \frac{m_X^2}{\mu^2} + \frac{1}{8} \right)
\nonumber \\
C_{S1}^{LR}(\mu) &=& -\frac{3}{2} \frac{\alpha_s}{4\pi},~~~~~~ 
C_{S2}^{LR}(\mu) 
                = 1- \frac{\alpha_s}{4\pi}, 
\ee
and the matrix elements estimated to be
\be
\langle O_1^{SLL}(1~{\rm TeV}) \rangle = -0.17,~~~
&&\langle O_2^{SLL}(1~{\rm TeV}) \rangle = -0.33,~~~~
\nonumber \\
\langle O_1^{LR}(1~{\rm TeV}) \rangle = -0.37,~~~
&&\langle O_2^{LR}(1~{\rm TeV}) \rangle = 0.51,
\nonumber \\
\langle O_1^{SLL}(m_t) \rangle = -0.14,~~~
&&\langle O_2^{SLL}(m_t) \rangle = -0.29,~~~~
\nonumber \\
\langle O_1^{LR}(m_t) \rangle = -0.30,~~~
&&\langle O_2^{LR}(m_t) \rangle = 0.40,
\ee
in (GeV)$^3$. 
We note that $C_i^{SLL}=C_i^{SRR}$, 
$\langle O_1^{SLL} \rangle=\langle O_1^{SRR} \rangle$.
The mass scale is taken to be 
$\mu = m_t(m_t)$ if $m_{H,A} < 1$ TeV and
$\mu = 1$ TeV elsewhere.


Figure 4 show the predictions of 
the  branching ratio ${\rm Br}(B_s \to \mu \tau)$
with respect to $m_H$
with allowed values of parameters given in the previous plots.
We find that the decay rates are substantial
and even there exists a lower limit of the branching ratio,
$\sim 3.5 \times 10^{-8}$.
These sizable $B_s \to \mu \tau$ decay rates
are caused by the $H$ exchange contribution.
Contributions of the CP-odd scalar $A$ also plays a role 
for these decay channels
since it cancels the $H$ contribution in $\Delta a_\mu$ 
but constructive in the $B_s$ decay rates. 

Observation of the LFV $B_s \to \mu \tau$ decays is a very clear
evidence of the NP, independent of the $h \to \mu \tau$ decays.
The detection of $\tau$ is still challenging at the LHC,
but the LHCb collaboration has reported the search results
of $B_s \to \tau^+ \tau^-$ and $B_d \to \tau^+ \tau^-$
with the $\tau$ reconstruction through the 3 prong decay
$\tau^- \to \pi^- \pi^+ \pi^- \nu_\tau$
\cite{Bstautau}.
Therefore we expect that it will be possible 
to observe $B_s \to \mu \tau$ decays in the future 
by achieving an improvement of the $\tau$ identification 
and more data sample in the experiment.

\begin{figure}
\begin{center}
\includegraphics[width=.6\textwidth]{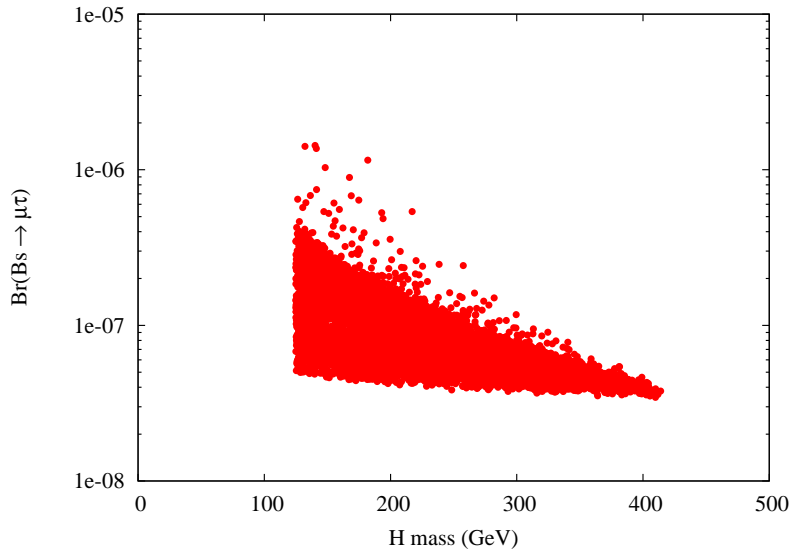}
\caption{
Branching ratios of $B_s \to \mu \tau$ decays with respect to $m_H$
which explain $h \to \mu \tau$ decays and
allowed by $(g-2)_\mu$, $\tau \to \mu \gamma$,
and $\tau \to \mu \mu \mu$ decays.
}
\label{fig:f4}
\end{center}
\end{figure}

\section{Concluding Remarks}
\label{sec:concl}

Inspired by the recent measurements of LFV $h \to \mu \tau$ decays,
we suggest an forbidden LFV $B_s$ decays into $\mu \tau $ final states
as a new signature of the LFV scalar interactions 
in the general 2HDM.
In order to accommodate the $\Delta a_\mu$ 
with the scalar FCNC in this model,
sizable contributions of additional scalars 
other than the SM Higgs boson are required.
We find that the scalar FCNC contributions to $\Delta a_\mu$
also induce large contribution to $B_s \to \mu \tau$ decays
and the considerable decay rates are possible.
We show that the branching ratio is larger than ${\cal O}(10^{-8})$
and even could be of order $\sim 10^{-5}$.
Such a large decay rate is possible to be measured at the LHCb
if $\tau$ detection is improved.

The scalar FCNC couplings in the quark sector, 
$h_{bs}$ are also essential to $B_s \to \mu \tau$ decays 
and constrained by the $B_s-\bar{B}_s$ mixing data.
The $bs$ FCNC couplings also lead to 
the NP contribution to $B_s \to \mu^- \mu^+$ decays in general,
of which recent measurement agrees with the SM prediction.
However our assumption of real $h_{bs}=h_{sb}$
makes NP contributions proportional to $h_{bs}$ and $h_{sb}$
cancel each other and thus
we consider no limits from $B_s \to \mu^- \mu^+$ decays
in this work.

The CMS and ATLAS results on $h \to \mu \tau$ have assumed
that the background is of the SM only and $m_h=125$ GeV.
The present signal strengths of the SM Higgs boson have errors
of order ${\cal O}(10~\%)$
\cite{PDG}. 
Thus they are not affected by the new decay channel to $\mu \tau$ 
of order 1 $\%$ branching fraction.
The mixing of $h$ and $H$ alters 
the SM Higgs couplings by the factor $\cos \theta$ in our model 
and also the signal strengths by $\cos^2 \theta$.
Since $|\sin \theta| < 0.07$, it is safe to assume
the SM background-only hypothesis.
Finally we consider the new scalar productions at the LHC.
In our analysis, the new scalar $H$ is not so heavy and
even less than 200 GeV, which is enough to be produced at the LHC.
However its ordinary Yukawa couplings are suppressed by $\sin \theta$
and the additional $h_ij$ couplings is assumed to be small. 
Therefore we do not worry about the LHC search bound 
on the new scalar bosons.

\section*{Acknowledgements}
KYL is supported by Basic Science Research Program
through the National Research Foundation of Korea (NRF)
funded by the Ministry of Science, ICT and Future Planning
(Grant No. NRF-2015R1A2A2A01004532).

\def\PRD #1 #2 #3 {Phys. Rev. D {\bf#1},\ #2 (#3)}
\def\NPRD #1 #2 #3 #4 {Phys. Rev. D {\bf#1},\ no.\ #2,\ #3 (#4)}
\def\PRL #1 #2 #3 {Phys. Rev. Lett. {\bf#1},\ #2 (#3)}
\def\PLB #1 #2 #3 {Phys. Lett. B {\bf#1},\ #2 (#3)}
\def\NPB #1 #2 #3 {Nucl. Phys. B {\bf #1},\ #2 (#3)}
\def\ZPC #1 #2 #3 {Z. Phys. C {\bf#1},\ #2 (#3)}
\def\EPJ #1 #2 #3 {Euro. Phys. J. C {\bf#1},\ #2 (#3)}
\def\JPG #1 #2 #3 {J. Phys. G: Nucl. Part. Phys. {\bf#1},\ #2 (#3)}
\def\JHEP #1 #2 #3 {JHEP {\bf#1},\ #2 (#3)}
\def\IJMP #1 #2 #3 {Int. J. Mod. Phys. A {\bf#1},\ #2 (#3)}
\def\MPL #1 #2 #3 {Mod. Phys. Lett. A {\bf#1},\ #2 (#3)}
\def\PTP #1 #2 #3 {Prog. Theor. Phys. {\bf#1},\ #2 (#3)}
\def\PR #1 #2 #3 {Phys. Rep. {\bf#1},\ #2 (#3)}
\def\RMP #1 #2 #3 {Rev. Mod. Phys. {\bf#1},\ #2 (#3)}
\def\PRold #1 #2 #3 {Phys. Rev. {\bf#1},\ #2 (#3)}
\def\IBID #1 #2 #3 {{\it ibid.} {\bf#1},\ #2 (#3)}

\end{document}